\documentclass[12pt]{iopart}

\usepackage{graphicx}
\usepackage{amssymb}
\usepackage{color}

\begin{document}

\title[Rotated multifractal network generator]{Rotated multifractal network generator}

\author{Gergely Palla$^1$, P{\'e}ter Pollner$^1$ and Tam\'as Vicsek$^{1,2}$ }

\address{$^1$ Statistical and Biological Physics Research Group of HAS and}
\address{$^2$ Dept. of Biological Physics, E{\"o}tv{\"o}s Univ.,
  1117 Budapest, P\'azm\'any P. stny. 1A}

\begin{abstract}
The recently introduced multifractal network generator (MFNG), has
 been shown to provide a simple and flexible tool for creating
random graphs with very diverse features. The MFNG is based
on multifractal measures embedded in 2d, leading also to isolated nodes,
whose number is relatively low for realistic cases, but may become dominant 
in the limiting case of infinitely large network sizes. Here we discuss the 
relation between this effect and the information dimension for the 
1d projection of the link probability measure (LPM), and argue that the node 
isolation can be avoided by a simple transformation of the LPM 
based on rotation.
\end{abstract}

\section{Introduction}
The network approach for describing complex natural, social and technological
phenomena has become very popular in the recent years. This can
be accounted for the generality of its fundamental concept of 
representing the connections among the units (building blocks) of the 
system under study with a graph \cite{Laci_revmod,Dorog_book}. Over the last
decade it has turned out that  networks corresponding to realistic systems
can be highly non-trivial, characterised by  a low average distance
combined with a high average clustering coefficient \cite{Watts-Strogatz},
anomalous degree distributions \cite{Faloutsos,Laci_science} and an intricate
modular structure \cite{GN-pnas,CPM_nature,Fortunato_report}. 

From the beginning of this new interdisciplinary field, {\it network models}
 have been playing a crucial role since they enable singling out 
the simplest aspects of complex structures and, thus, are extremely useful
in understanding the underlying principles. Furthermore, models
can also help testing hypotheses about measured data. In parallel with 
the discovery of the fine structure of real networks, many important 
and successful models have been introduced over the past 10 years
for interpreting the different aspects of the studied systems. 
However, most of these models explain only a particular aspect of
the network (clustering, a given degree distribution, etc.), and for
each newly discovered feature a new model had to be constructed.

Due to this proliferation of network models, the concept of general 
network models and methods for generating graphs with desired properties
has attracted great interest lately. A number of noteworthy methods have
been proposed starting from the exponential random graph 
model \cite{Frank-exp,Wasserman-exp,snijders-exp,Newman_sol}, 
through the hidden variable models
 \cite{caldarelli_fitt,hidden_vars} (including the systematic study of the
entropy of network ensembles \cite{Bianconi}), the $dK$ series approach
\cite{dk-series} and the use of $p$-adic randomised Parisi matrices 
\cite{Avetisov,Avetisov_new} to the Kronecker-graph approach 
\cite{Leskovec_elso,Leskovec_two}. 

A very recently introduced approach
along this line is the multifractal network generator (MFNG) \cite{our_PNAS},
which was shown to be capable of  generating a wide variety of network types 
with prescribed statistical properties, 
(e.g., with degree- or clustering coefficient 
distributions of various, very different forms). At the heart of this method
lies a mapping between 2d measures defined on the unit square and random
graphs. The main idea is to iterate a suitably chosen self similar 
multifractal (becoming singular in the limiting case) and enlarge the
 size of the generated graph (becoming infinite in the limiting case)
in parallel. A very unique feature of this construction is that with the 
increasing system size the generated graphs become topologically more 
structured. 

However, a slight drawback of the method is that when the size of the
generated networks (and in parallel, the number of iterations in
the multifractal) grow to infinity, isolated nodes overtake the majority 
of the graph in most settings, 
(see the SI of Ref.\cite{our_PNAS} for more details). 
Although this effect is usually not an issue when constructing graphs of sizes
comparable to real networks, finding a way to circumvent it would still 
provide a noteworthy improvement, especially in the light of the 
non-trivial connections between convergent graph sequences in the 
infinite network size limit and 2d functions on the unit square \cite{lovasz-szegedy,lovasz_cikk}. 

In this article we study the relation between the node isolation effect
and the 1d projection of the link probability measure (LPM) used in the
graph generation process. Furthermore, we propose a natural method to
overcome the problem by a simple transformation based on rotation. The
paper is organised as follows. In Sect.\ref{sect:mfng}. we overview
the definition and most important properties of the 
multifractal network generator, while in Sect.\ref{sect:multifract}
we discuss the connection between node isolation and multifractality. 
We continue by proposing a modification of the original method
avoiding the node isolation in Sect.\ref{sect:rot}., which is tested
 in practise in Sect.\ref{sect:app}. 
Finally, we conclude in Sect.\ref{sect:sum}.

\section{The multifractal network generator}
\label{sect:mfng}

The multifractal network generator was inspired by earlier results
from L. Lov\'asz and co-workers proving that in the infinite network 
size limit, a dense graph's adjacency matrix can be well represented by 
a continuous function $W(x,y)$ on the 
unit square \cite{lovasz-szegedy,lovasz_cikk}. 
A similar approach was introduced by Bollob\'as et al.
in Refs.\cite{Bollobas-cikk,Bollobas_chapter} and
was used to obtain convergence and phase transition results for
inhomogeneous random (including sparse) graphs. This two variable
symmetric function
 (which can have a very simple form for a variety of interesting
  graphs, and was supposed to be either continuous or almost everywhere 
continuous) predicts the probability whether two nodes are
 connected or not. 

\begin{figure}[hbt]
\centerline{\includegraphics[width=0.8\textwidth]{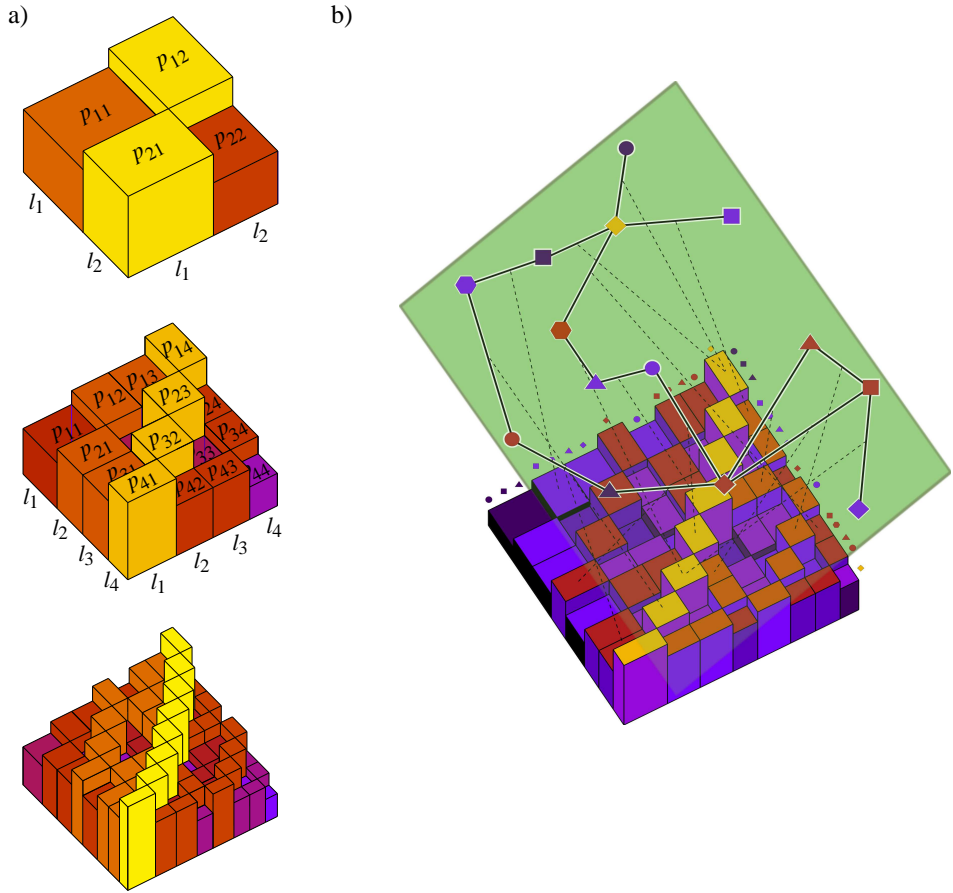}}
\caption{Schematic illustration of the multifractal graph generator. a) 
The construction of the link probability measure. 
We start from a symmetric generating measure on the unit square defined by a
 set of probabilities $p_{ij}= p_{ji}$ associated with $m\times m$ 
rectangles (shown on the left). The generating measure is iterated by 
recursively multiplying each box with the generating measure itself as shown
in the centre and on the right. b) Drawing linking probabilities from the obtained LPM. We assign random coordinates in the unit interval to the nodes in
the graph and link each node pair $I$, $J$ with a probability given by the 
probability measure at the corresponding coordinates, resulting in a 
graph drawn in the transparent green layer above the LPM.}
\label{fig:mfng_illustr}
\end{figure}

In case of the MFNG the
mentioned $W(x,y)$ is replaced by a
 self-similar multifractal \cite{our_PNAS}. 
We start by defining a generating measure on the unit
 square by dividing identically both the $x$ and $y$ axis to $m$ 
(not necessarily equal) intervals, splitting it to $m^2$
rectangles, and assigning a probability $p_{ij}$ to each rectangle
($i,j\in[1,m]$ denote the row and column indices). The probabilities
are assumed to be normalised, $\sum p_{ij}=1$ and symmetric $p_{ij}=p_{ji}$.
(Note that we normalize the probabilities of the generating measure
instead of the integral of the latter because of the advantages of
this choice to be discussed later.)
 Next, the LPM is obtained by recursively multiplying
 each rectangle with the generating measure $k$ times. 
This is in complete analogy with the standard
process of generating a multifractal, resulting in
$m^{2k}$ rectangles, each associated with a linking probability
 $p_{ij}(k)$ equivalent to a product of $k$ factors from the
original generating $p_{ij}$ given as
\begin{equation}
p_{ij}(k)=\prod_{q=1}^{k}p_{i_{q}j_{q}}.
\label{eq:p_ij_prod}
\end{equation}
 In our convention $k=1$ stands for
the generating measure, thus, an LPM at $k=1$ is equivalent to the generating measure itself. The indices of the factors in
(\ref{eq:p_ij_prod}) are given by
\begin{equation}
i_{q}=\left\lfloor\frac{(i-1)\prod_{r=1}^{q-1} \circ \,{\rm mod}\, m^{k-r}}{m^{k-q}}\right\rfloor+1,
\label{eq:indices}
\end{equation}
where $\lfloor a/b\rfloor$ denotes the quotient (integer part) of $a/b$,
the term $\prod_{r=1}^{q-1} \circ \,{\rm mod}\, m^{k-r}$ stands for subsequent
calculation of the remainder after the division by $m^{k-r}$, and an
 analogous formula can be written for the indices $j_q$ as well.
(For $q=1$, Eq.(\ref{eq:indices}) simplifies to
$i_{q}=\lfloor (i-1)/m^{k-1}\rfloor+1$). To obtain a network from the 
link probability measure, we distribute 
$N$ points independently,
 uniformly at random on the $[0,1]$ interval, and link each pair
with a probability given by the $p_{ij}(k)$ at the given coordinates. 
(The above process is illustrated in Fig.\ref{fig:mfng_illustr}).

The diversity of the linking probabilities $p_{ij}(k)$ (and correspondingly,
the structuredness of the generated graph) 
is increasing with the number
of iterations, just like in case of a standard multifractal. When considering
the ``thermodynamic limit'' of this construction 
($k\rightarrow\infty$, $N\rightarrow\infty$) we would like to keep
the generated networks sparse, i.e., ensure that the average degree
of the nodes, $\left< d\right>$ remains constant. This can be achieved by 
an appropriate choice of the number of nodes
as a function of $k$, using the following relation:
\begin{equation}
\left< d\right>=N(k)\sum_{i=1}^{m^k}\sum_{j=1}^{m^k}p_{ij}(k)a_{ij}(k),
\label{eq:av_d}
\end{equation}
where $a_{ij}(k)$ denotes the area of the box $i,j$ at iteration $k$.
For simplicity, let us consider the special case of equal sized boxes $a_{ij}(k)=m^{-2k}$. Due to
the normalisation of the linking probabilities in this case 
the above expression simplifies
to $\left< d\right>=Nm^{-2k}$, thus, to keep the average degree constant
when increasing the number of iterations for a given generating measure, the
 number of nodes have to be increased exponentially with $k$.

The above construction could be made more general by replacing
the ``standard'' multifractal with the $k$-th tensorial product of a 
symmetric 2d function $0\leq W(x,y)\leq 1$ defined on the unit square. 
Although the resulting 
$W_k(x_1,...,x_k,y_1,...,y_k)=W(x_1,y_1)\cdots W(x_k,y_k)$ 
function is $[0,1]^{2k}\rightarrow[0,1]$ instead of $[0,1]^2\rightarrow[0,1]$,
with the help of a measure preserving bijection between $[0,1]$ and
 $[0,1]^{k}$ it could be used to generate random graphs in the same 
manner as with $p_{ij}(k)$.

At this point we also note that omitting the normalisation condition 
$\sum_{ij}p_{ij}=1$ for the generating measure gives the approach an additional
flexibility which can come very handy in practical cases. Suppose that for a 
given setting of $k$, $N$ and the LPM we would like to increase the average 
degree in the obtained graph beside preserving the relative ratios of the
linking probabilities (expected degrees) of nodes falling into the different
rows of the LPM. A very natural idea in this case is to multiply each 
element in $p_{ij}(k)$ with the same factor $\eta>1$, and use the resulting
matrix for generating a random graph in the same way. However, this 
multiplicative factor could be also introduced at the level of the generating 
measure instead, i.e., $\eta^{1/k}p_{ij}$ would also generate $\eta p_{ij}(k)$
for the LPM. 

\subsection{The degree distribution}
\label{sect:mfng_deg}

An important property of the MFNG is that nodes with coordinates falling 
into the same row (column) of the
LPM are statistically identical. This means
that e.g., the expected degree or clustering coefficient of the
nodes in a given row is the same. Consequently, the distributions
related to the topology are composed of sub-distributions associated
with the individual rows. The degree distribution can
be expressed as
\begin{equation}
\rho^{(k)}(d)=\sum_{i=1}^{m^k}\rho_i^{(k)}(d)l_i(k),
\label{eq:deg_dist_alap}
\end{equation}
where $\rho_i^{(k)}(d)$ denotes the sub-distribution of the nodes in
row $i$, and $l_i(k)$ corresponds to the width of the row
(giving the ratio of nodes in row $i$ compared to the number of
total nodes). These $\rho_i^{(k)}$ take the form of \cite{our_PNAS}
\begin{equation}
\rho_i^{(k)}(d)=\frac{\left< d_i(k)\right>^{d}}{d!}e^{-\left< d_i(k)\right>},
\label{eq:row_deg_dist}
\end{equation}
where $\left< d_i(k)\right>$ denotes the average degree 
of nodes in row $i$. This is is given by $\left< d_i(k)\right>=N(k) p_i(k)$,
where $p_i(k)$ corresponds to the linking probability in row $i$, given by
\begin{equation}
p_i(k)=\sum_j p_{ij}(k)l_j(k)
\label{eq:p_i}
\end{equation}
In more general, if the linking probability at iteration $k$ is described
by $W_k(x,y)$, then the expected degree of a node having a position 
$x$ can be given as
\begin{equation}
\left< d(x)\right>=N(k) w_k(x),
\label{eq:d_exp}
\end{equation}
where 
\begin{equation}
w_k(x)\equiv \int dy W_k(x,y)
\label{eq:w_kx}
\end{equation}
defines the 1d projection of $W_k(x,y)$ and is equivalent to the linking
probability at position $x$. In case of the multifractal
 network generator this $w_k(x)$ has a simple step-wise constant form,
 showing a step-wise surface getting rougher and rougher with increasing
$k$.

\section{The isolated nodes and the multifractality of $w_k(x)$}
\label{sect:multifract}

According to Sect.\ref{sect:mfng_deg}., the degree distribution and the fraction of 
isolated nodes depend on the projection of the link probability measure
 $p_{ij}(k)$ to the 
$x$ axis, (or equivalently, to the $y$ axis), 
given by $w_k(x)$. It is known that almost all projections of a 
multifractal such as $p_{ij}(k)$ with an information dimension larger than 1 
to a 1d line result in measures
with Euclidean support \cite{projection}. 
However, the projection to $w_k(x)$ is unfortunately a special
case, belonging to the minority of projections yielding the 
``almost all'' instead of ``every'' in the previous statement. 
As we shall see shortly, $w_k(x)$ is a multifractal itself with an 
information dimension smaller than 1, and this is the origin of the 
node isolation.

\begin{figure}[hbt]
\centerline{\includegraphics[width=0.8\textwidth]{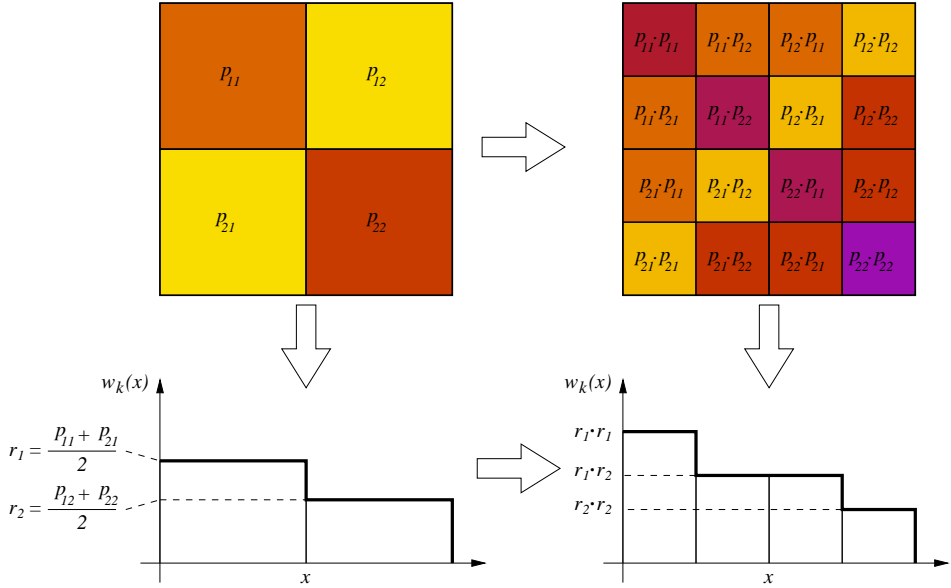}}
\caption{
Projection of the LPM onto the edge of the 
unit square, resulting in a multifractal $w_k(x)$ function. For simplicity
we assumed a 2 by 2 generating measure with equal box lengths.
}
\label{fig:proj_illustr}
\end{figure}

In Fig.\ref{fig:proj_illustr}. we show $w_k(x)$ for $k=1$ and $k=2$ in a
setting with equal sized boxes in the generating measure. For simplicity
we shall assume equal sized boxes in the rest of this Section.
 Since the linking probability inside each box is constant, the shape of
of $w_k(x)$ is step-wise, consisting of $m^k$ intervals (corresponding
to the columns of the LPM), and the height 
of step $i$ is given by 
\begin{equation}
r_i(k)\equiv\sum_{j=1}^{m^k}p_{ij}(k)l_{j}(k).
\label{eq:r_ik}
\end{equation}
 However, the multiplicative nature
of the construction is inherited by $w_k(x)$ as well,
\begin{equation}
r_i(k)=\prod_{q=1}^kr_{i_q},
\end{equation}
where 
\begin{equation}
r_i\equiv \sum_{j=1}^mp_{ij}l_j,
\label{eq:r_i}
\end{equation}
stand for the heights of the steps at the generating measure ($k=1$), and
$i_q$ is given by (\ref{eq:indices}). (This is demonstrated in 
Fig.\ref{fig:proj_illustr}. for $k=2$).  Thus, the evolution
of $w_k(x)$ with $k$ is analogous to the standard construction of 
a multifractal embedded in the unit interval. Note however that 
if $\sum_{ij}p_{ij}=1$, then $w_k(x)$ is not normalised,
e.g., for equal box lengths $\int w_k(x) dx=m^{-2k}$.

Multifractals are described by the $q$ ordered generalised 
fractal dimension $D(q)$ defined as follows (see e.g., Ref.\cite{Tamas_konyv}).
 Suppose that 
we divide the multifractal to boxes of size $\epsilon$, and the measure 
inside box $i$ is given by $p_i$. The function $\chi(q,\epsilon)$ is defined as 
\begin{equation}
\chi(q,\epsilon)\equiv \sum_i p_i^q.
\end{equation}
If $\epsilon$ is varied, $\chi(q,\epsilon)$ behaves as
\begin{equation}
\chi(q,\epsilon)\sim \epsilon^{D(q)(q-1)}.
\end{equation}
Thus, $D(q)$ can be given as
\begin{equation}
D(q)=\lim_{\epsilon\rightarrow 0}\left[\frac{1}{q-1}\frac{\ln\sum_ip_i^q}{\ln \epsilon}\right].
\label{eq:def_Dq}
\end{equation}
In the special case of $q=1$ we have zero in the denominator, thus
we take the $q\rightarrow 1$ limit and using l'Hospital's rule we obtain
\begin{equation}
-\sum_i p_i\ln p_i\sim D_1\ln(1/\epsilon).
\end{equation}
From the point of view of the degree distribution and the fraction
of isolated nodes, the crucial $q$ value is $q=1$: When the number of 
iterations, $k\rightarrow\infty$, the fractal dimension of
the support of the measure is given by $D(q=1)$. If it turns
out that $D(q=1)<1$ for the $w_k(x)$ curve, then this means that
the points giving the 
relevant contribution to the occurrence of links are concentrated 
on a fractal with a fractal dimension
smaller than one, and thus, the majority of the nodes become isolated.

For a multifractal on the unit interval defined by a self-similar 
multiplication process such as in case of $w_k(x)$, (governed by 
Eqs.(\ref{eq:r_ik}-\ref{eq:r_i})),  the $D(q)$ can be calculated 
analytically \cite{Tamas_konyv} as
\begin{equation}
D(q)=\frac{1}{(q-1)\ln(1/m)}\ln\left[\sum_{i=1}^m(r_im)^q\right].
\label{eq:D_q_simp}
\end{equation}
For $q=0$ the above expression yields $D(q=0)=1$, in agreement
with the general picture of multifractals produced in a 
recursive multiplication process, where $D(q=0)$ equals the 
fractal dimension corresponding to a uniform generating object.
For any multifractal in general, the $D(q)$ values monotonically decrease
with increasing $q$. Thus, at $q=1$ in our case the $D(q)$ can reach $1$ 
only if $D(q)=1$ for any $q\in[0,1]$.
According to (\ref{eq:D_q_simp}),
 this can be achieved only if $r_i=1/m^2$ for 
all $i$. This means that unless the sum of probabilities in any row of the
generating measure is the same, the $D(q=1)$ becomes smaller than 1, 
and the node isolation effect takes place.

\section{Rotated measures}
\label{sect:rot}

\begin{figure}[h]
\centerline{\includegraphics[width=0.65\textwidth]{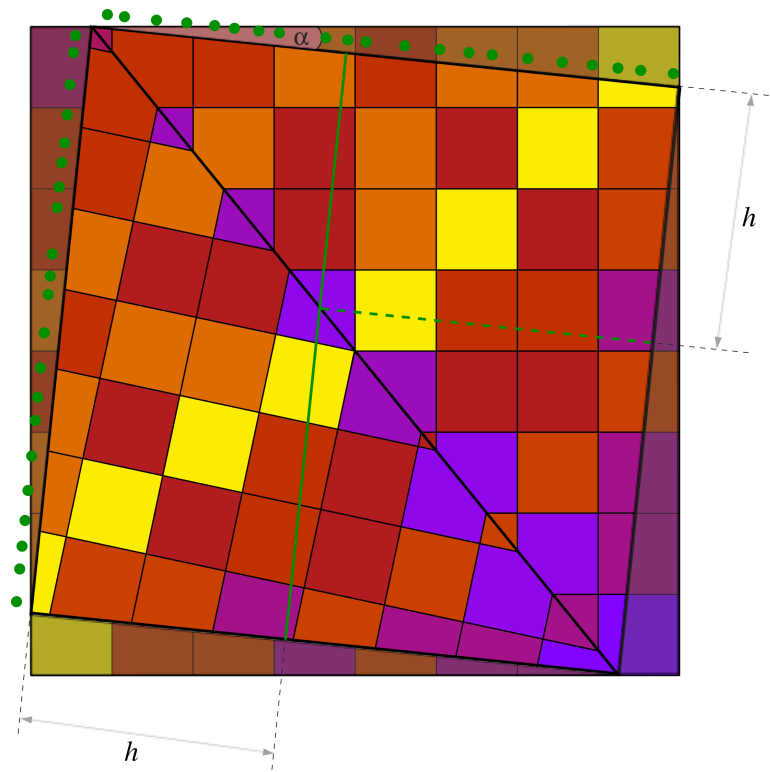}}
\caption{A ``standard'' link-probability measure (oblique) with a 
rotated frame inside. The probabilities inside the rotated square
have to be symmetrised along the diagonal. The linking probability of the
green node can be calculated by summing up the probabilities ``traversed''
by the green line, each multiplied by the length of the intersection between
the green line and the corresponding box. Due to the symmetry, this is
equivalent to traversing first along the original line from the top 
until the diagonal is met, and then continue along the dashed line to 
the right.}
\label{fig:rot_frame}
\end{figure}

Our aim is to overcome the problem of the exact multifractality of $w_k(x)$ 
by modifying the construction in such a way that the projection of
$W_k(x,y)$ determining the degree distribution has Euclidean support.
Meanwhile, we want to keep the LPM a highly variable function so that
very different distributions (leading to very different kinds of networks) 
could be still achieved.
A simple idea is to rotate the LPM with
a given angle $\alpha$ as shown in Fig.\ref{fig:rot_frame}., so that
the direction of the projection determining the degree distribution 
no longer coincides with any of
the special directions of the multifractal generation process. 
Thus, the construction of a random graph in this setting has
the following main stages: we begin by generating a ``standard'' LPM
 for a chosen $k$, and ``cut'' a square rotated by an angle $\alpha$ 
from this measure as shown in Fig.\ref{fig:rot_frame}. Since the diagonal
of this newly introduced square does not coincide with the diagonal of the
original LPM, we have to symmetrise the measure inside the rotated square
with respect to its diagonal. Finally,  we distribute
$N$ points uniformly at random along both sides of the rotated 
square and link each pair with a probability found at the given coordinates
in the ``rotated coordinate system''. (This last step is in complete 
analogy with the ``standard'' graph generation process).

When examining the behaviour of this construction with increasing $k$, 
the number of nodes used in the graph generation is 
adjusted according to the $\left< d\right>=$const. criterion,
just as in case of the original settings. 
Due to the rotation and the symmetrisation, the polygons making
up the link probability matrix are no longer arranged in a matrix like
form, thus, (\ref{eq:av_d}) cannot be used straight away in this case. 
To calculate $\left< d\right>$ we need to first introduce a unique indexing
over the polygons inside the rotated frame, and evaluate 
\begin{equation}
\left< d\right> = N\sum_i p_ia_i,
\label{eq:rot_av_d}
\end{equation} 
where $p_i$ and $a_i$ denote the probability and the area of the polygon $i$
 (triangles, quadrangles and pentagons) defining the rotated link 
probability measure. The size of the various distances and the 
equations for the most important lines needed to evaluate 
(\ref{eq:rot_av_d}) are given as a function of 
the rotation angle $\alpha$ in the Appendix.

The expected degree of a node at distance $h$ from the 
origin of the rotated square can be given by an expressions analogous to
(\ref{eq:d_exp}) as 
\begin{equation}
\left< d(h)\right> = N(k)w_k(h),
\end{equation}
where the linking probability $w_k(h)$ has to be calculated along a line
parallel to the side of the rotated square. By denoting the set of
polygons intersected by this line by $\Omega_k(h)$, we can express
$w_k(h)$ as
\begin{equation}
w_k(h)=\sum_{j\in\Omega_k(h)}p_j\tilde{l}_j(h),
\label{eq:w_k_h}
\end{equation}
where  $\tilde{l}_j$ denotes the relative length of the intersections divided by
the length of the side of the rotated frame (see Fig.\ref{fig:rot_frame}).
(These intersection lengths can be calculated with simple coordinate
geometry, based on the equations given in the Appendix).
Due to the symmetrisation of the rotated square, the boxes under the diagonal
are rotated by $2\alpha$. Thus, in practise it is more simple to 
replace the summation (\ref{eq:w_k_h}) 
along a straight line by a summation along a broken line which is fully 
above the diagonal, as shown by the dashed green line in 
Fig.\ref{fig:rot_frame}. 

\begin{figure}[hbt]
\centerline{\includegraphics[width=0.65\textwidth]{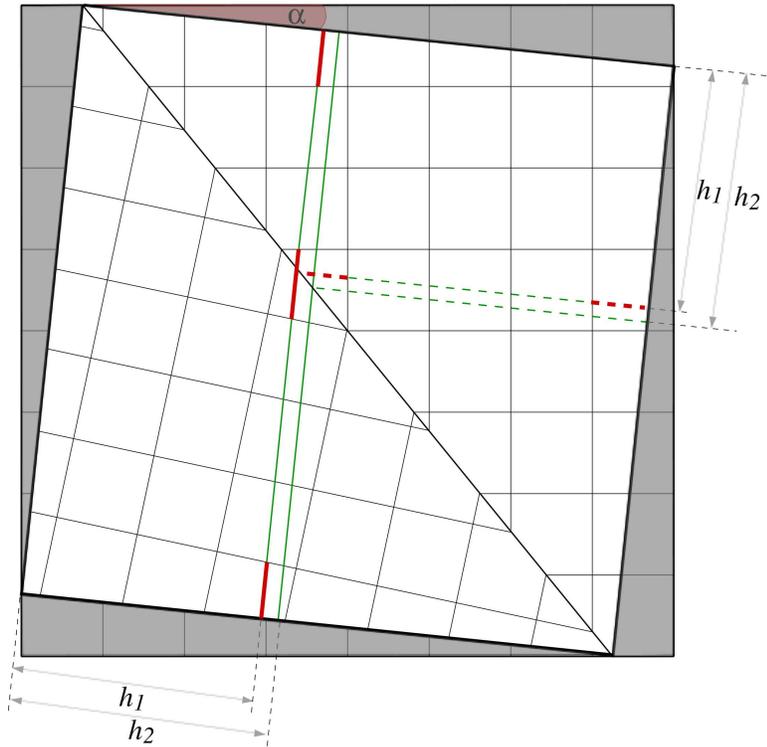}}
\caption{Two node positions in the rotated square, relatively close to
each other shown by the green lines. 
They intersect with the same boxes, and in most cases the
length of the intersections is the same as well. The different intersection
lengths are marked by red.}
\label{fig:h_idea}
\end{figure}

 The $w_k(h)$ function
is the analogue of the $w_k(x)$ function for the rotated frame, and
when $\alpha\rightarrow0$, $w_k(h)\rightarrow w_k(x)$.
As already noted in Sect.\ref{sect:mfng}, the 1d projection of
the original link probability measure, $w_k(x)$, is always
piecewise constant, where the constant intervals correspond to the
columns of the link-probability measure. In case of $w_k(h)$ 
the situation is a bit more complex. In Fig.\ref{fig:h_idea}. we depict
two close by node positions in the rotated square. The lines along which one
has to calculate the linking probabilities intersect with the same
boxes. Furthermore, the lengths of the intersections are the same for the 
two lines in most of the boxes, except for the sections marked by red. Thus,
the linking probability of the two nodes given by $w_k(h_1)$ and $w_k(h_2)$ 
will be quite close to each other
as well, with the difference coming from the few different intersection lengths.
Now let us imagine that we fix one of these two node positions, 
and set the corresponding
linking probability $w_k(h)$ as a reference value. If we scan with the other 
node through a narrow interval of $h$ values such that the corresponding
line still intersects
with the same boxes, then due to the linear change in the intersection
lengths, the change in $w_k(h)$  with respect to the reference value 
will be linear as well as a function of the $h$ difference between 
the two node positions. 

From this it
follows that if we scan through the total range of possible $h$ values, 
then the corresponding $w_k(h)$ curve of the 
linking probabilities will be \emph{piece wise linear}. 
The break points between the linear segments
correspond to the $h$ values where the line parallel to the
rotated square boundary comes across a corner
of the polygons defining the 2d link probability measure, and starts to 
intersect with a new polygon. Thus, to analytically calculate $w_k(h)$ we
 need to evaluate (\ref{eq:w_k_h}) only at the break points, and connect
the results with linear segments.
The $h$ value of the break points (corresponding to polygon corners) can
be most easily calculated by changing the coordinate system to be aligned
with the rotated square, having an origin at the top left corner.
The details of this coordinate transformation are given in the
 Appendix. In Fig.\ref{fig:h_prof_check}a we check the above for a 
2 by 2 generating measure at $k=3$ by calculating $w_k(h)$ in both the
break points and in two intermediate points between each adjacent break
point pairs. Seemingly, the results for the intermediate points fall on
the lines connecting the result for the break points.

The degree distribution can be obtained from $w_k(h)$ in three simple steps.
 The first step is the calculation of 
the distribution of the linking probability for the nodes, $\sigma^{(k)}(p)$.
 (By integrating $\sigma^{(k)}(p)$ as $\int_{p_1}^{p_2}\sigma^{(k)}(p)dp$
we receive the probability for a randomly chosen node to obtain a linking
probability between $p_1$ and $p_2$, and the number of expected 
links on the node is given by the total number of nodes, $N$, multiplied by
its linking probability). The $\sigma^{(k)}(p)$ can be calculated from 
$w_k(h)$  by a simple ``projection'' to the vertical axis as follows. 
Since we distribute the 
nodes uniformly at random along the side of the rotated square, the ratio
of nodes falling into an interval of $[h_1,h_2]$ is simply $(h_2-h_1)/b$,
where $b$ denotes the length of the side of the rotated square. Thus,
a linear segment of $w_k(h)$, stretching from $h_1$ to $h_2$ contributes
to $\sigma^{(k)}(p)$ with a ``step'' ranging from $p_1=w_k(h_1)$ to 
$p_2=w_k(h_2)$ with a height of $(h_2-h_1)/(b|p_2-p_1|)$. 
(For an illustration see Fig.\ref{fig:h_prof_check}b).

The second step in the calculation of the degree distribution is the 
transformation of $\sigma^{(k)}(p)$ into the distribution of the 
expected degrees for the nodes, $\tilde{\rho}^{(k)}(d)$. The difference
between the degree distribution and $\tilde{\rho}^{(k)}(d)$ can be 
illustrated in the graph generation process: the expected degree
of a node is simply its linking probability given by (\ref{eq:w_k_h})
multiplied by $N$, however, since the links are drawn randomly, its actual
degree may become smaller or larger than that at the end of the 
link generation process. The $\tilde{\rho}^{(k)}(d)$ can be obtained 
from $\sigma^{(k)}(p)$ by a simple ``stretching'' in the horizontal direction,
 i.e., for any $p_1$ and $p_2$
\begin{equation}
\int_{p_1}^{p_2}\sigma^{(k)}(p)dp=\int_{x_1=Np_1}^{x_2=Np_2}\tilde{\rho}^{(k)}(x)dx,
\end{equation}
where the integral on the right hand side corresponds to the probability
for a randomly chosen node to have an expected degree falling between
$d_1=Np_1$ and $d_2=Np_2$. We note that in case of the original
settings without any rotation, both $\sigma^{(k)}(p)$ and 
$\tilde{\rho}^{(k)}(d)$ are given by trains of delta
 spikes with varying weights, 
\begin{eqnarray}
\sigma^{(k)}(p)&=&\sum_{i=1}^{m^k}l_i(k)\delta_{p,p_i(k)}, \\
{\tilde{\rho}}^{(k)}(d)&=&\sum_{i=1}^{m^k}l_i(k)\delta_{d,Np_i(k)},
\label{eq:orig_exp_ddist}
\end{eqnarray}
where $p_i(k)$ denotes the linking probability
in column $i$ of the original LPM given by 
(\ref{eq:p_i}). In contrast, for the rotated measures both 
$\sigma^{(k)}(p)$ and ${\tilde{\rho}}^{(k)}(d)$ take a step-wise
form instead of delta spikes, (see e.g., Fig.\ref{fig:h_prof_check}b). 

The final step is to transform ${\tilde{\rho}}^{(k)}(d)$ into the degree
 distribution. Since the links are drawn
independently of each other in both the original and the rotated
settings, this can be achieved 
by taking the convolution of ${\tilde{\rho}}^{(k)}(d)$ with a 
Poisson-distribution as
\begin{equation}
\rho^{(k)}(d)=\int dx {\tilde{\rho}}^{(k)}(x)\frac{x^{d}}{d!}e^{-x}.
\end{equation}
\begin{figure}[h]
\centerline{\includegraphics[width=\textwidth]{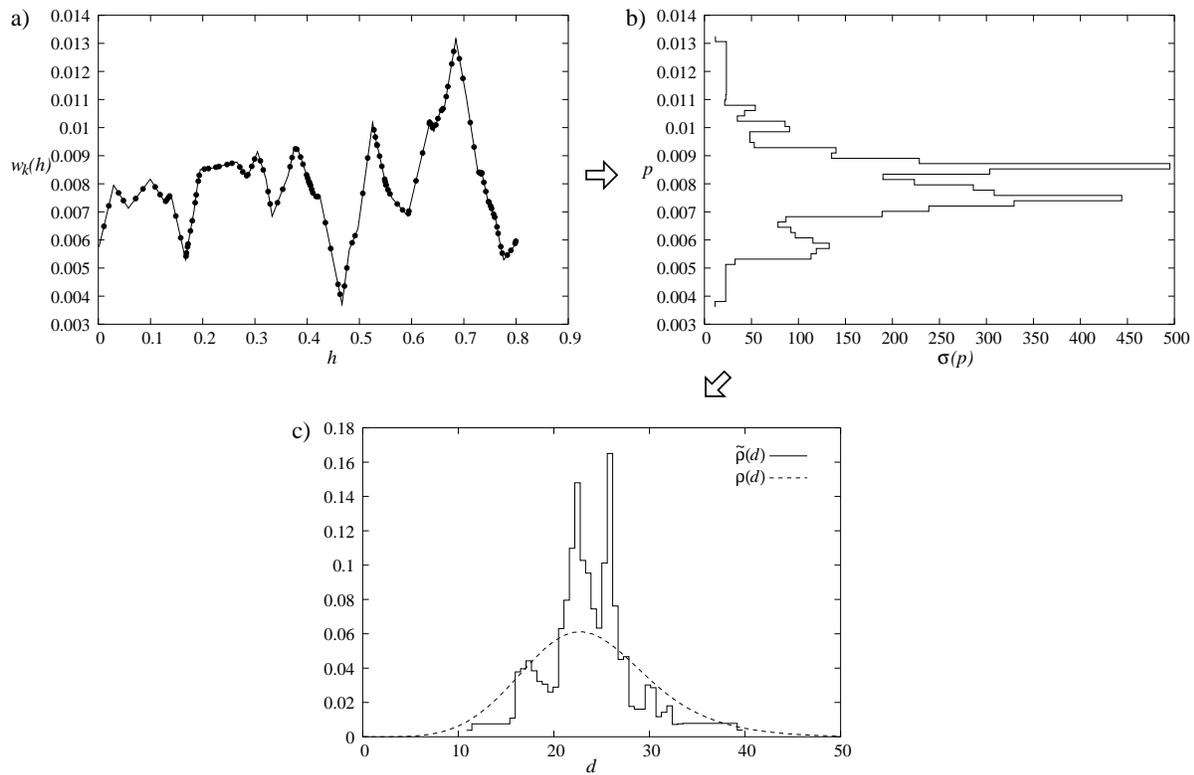}}
\caption{a) Checking the $w_k(h)$ of a 2 by 2 generating
measure at $k=3$. The continuous line shows the 
piece wise linear linking probability as a function of $h$ obtained by
connecting the values obtained at adjacent break points with straight lines.
For each section we also calculated the linking probability in two intermediate
points as well, the results are shown by the symbols. b) By projecting 
$w_k(h)$ to the vertical axis one obtains the distribution 
of the linking probabilities, $\sigma(p)$, which is step-wise. (Note that in
order to emphasise the connection to $w_k(h)$, we have plotted $p$ on the
vertical axis and $\sigma(p)$ on the horizontal axis). 
c) The distribution of the expected degrees, $\tilde{\rho}^{(k)}(d)$ is 
obtained by inflating $\sigma(p)$ according to the $\left< d\right>=Np$ 
relation. We have also plotted
 the corresponding degree distribution, $\rho^{(k)}(d)$, with dashed lines.}
\label{fig:h_prof_check}
\end{figure}
Based on the method detailed above, in the next Section we compare the 
evolution of the degree distribution with $k$ in the original settings and 
in the rotated scenario, (with a special focus on the ratio of the
isolated nodes).

\section{Applications}
\label{sect:app}

\subsection{Degree distribution}
\label{sect:deg}

\begin{figure}[hbt]
\centerline{\includegraphics[width=\textwidth]{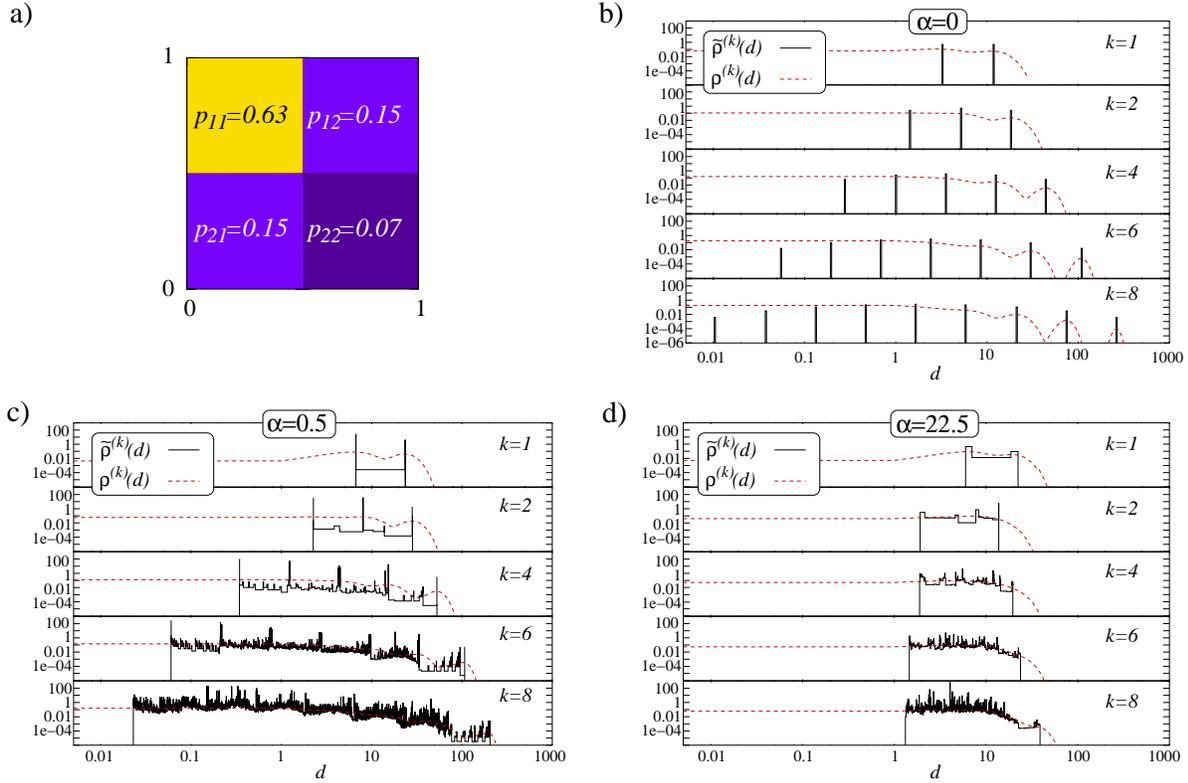}}
\caption{a) A 2 by 2 generating measure with equal box lengths. 
b) The distribution of the expected degrees,
${\tilde{\rho}}^{(k)}(d)$, (black, solid lines), and the degree distribution,
$\rho^{(k)}(d)$, (red, dashed lines) in
the original settings for $k=1,2,4,6,8$ on logarithmic scale. 
Since ${\tilde{\rho}}^{(k)}(d)$ corresponds to a delta-spike train according to 
(\ref{eq:orig_exp_ddist}), we used logarithmic binning for dealing with
the singularities.
c) The ${\tilde{\rho}}^{(k)}(d)$ (black, solid lines), and $\rho^{(k)}(d)$, 
(red, dashed lines) 
distributions at $k=1,2,4,6,8$ for a rotated frame at
rotation angle $\alpha=0.5$ degrees. d) The ${\tilde{\rho}}^{(k)}(d)$ 
(black, solid lines), and $\rho^{(k)}(d)$, 
(red, dashed lines) 
distributions at $k=1,2,4,6,8$ for a rotated frame at
rotation angle $\alpha=22.5$ degrees.}
\label{fig:rot_alfa_1}
\end{figure}

For the comparison between the original settings and the rotated
scenario we chose the 2 by 2
generating measure shown in Fig.\ref{fig:rot_alfa_1}a. with a 
starting $N_{k=1}=30$, since the
ratio of isolated nodes becomes significant quite fast without the
 rotation of the LPM in this case. The generating measure and 
$N_{k=1}$ fixed the average degree of the nodes to $\left< d \right>=7.5$. From $\left< d\right>$ we can calculate the
number of nodes at the higher $k$ values in the original setting
from (\ref{eq:av_d}). In case of the rotated scenario we use the
same $\left< d\right>$, and calculate the number of nodes for any $k$ value
from (\ref{eq:rot_av_d}). 

In Fig.\ref{fig:rot_alfa_1}b we show the distribution of the 
expected degrees for the nodes, ${\tilde{\rho}}^{(k)}(d)$, on
logarithmic scale for the 
original settings, obtained from (\ref{eq:orig_exp_ddist}), whereas 
Fig.\ref{fig:rot_alfa_1}c-d show the same function  for
rotated frames at rotation angles $\alpha=0.5$ degrees and $\alpha=22.5$
 degrees, respectively. We also plotted the corresponding degree distributions
 with dashed lines, however, the difference between the original and the 
rotated scenario is much more salient for ${\tilde{\rho}}^{(k)}(d)$.
 As explained in Sect.\ref{sect:rot}.,
${\tilde{\rho}}^{(k)}(d)$ consists of delta-spikes in case of the 
original settings (we have used binning in order to plot this 
singular function), and according to  Fig.\ref{fig:rot_alfa_1}b, as
$k$ is increased, the distribution gets wider, and a significant part
of it is shifted under $d=1$. This means that as we increase 
$k$, for larger and larger part of the nodes the expected degree 
becomes smaller than one,  thus, the node isolation effect takes
place. In contrast, Figs.\ref{fig:rot_alfa_1}c-d show a different
 behaviour. Although the distributions become wider with increasing 
$k$ here as well,
this tendency is much less pronounced compared to the $\alpha=0$ case.
 Furthermore, in case of Fig.\ref{fig:rot_alfa_1}d the major part of the
distribution stays above $d=1$ for the examined $k$ values.

\begin{figure}[hbt]
\centerline{\includegraphics[width=\textwidth]{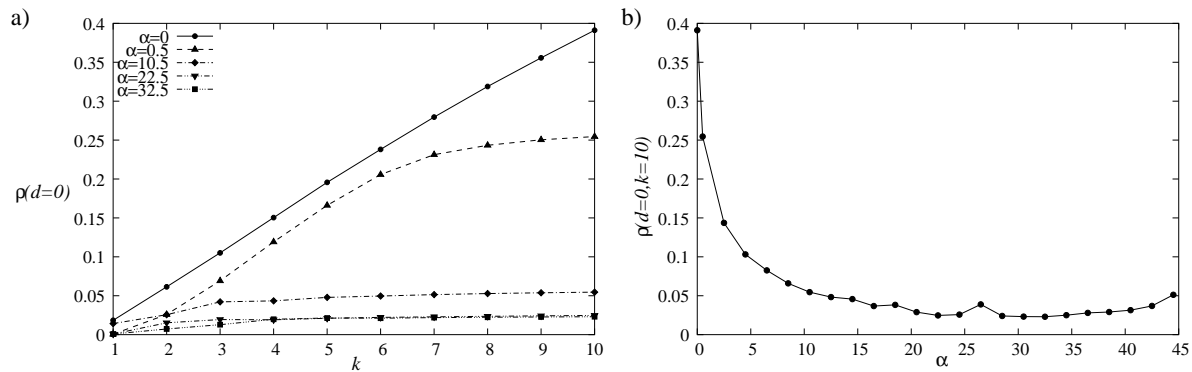}}
\caption{ a) The ratio of isolated nodes as a function of the number 
of iterations for the  examined generating measure 
(shown in Fig.\ref{fig:rot_alfa_1}a). b) The ratio of isolated nodes 
at $k=10$ as a function of the rotation angle $\alpha$. }
\label{fig:rot_alfa_1_deg_zeros}
\end{figure}

In Fig.\ref{fig:rot_alfa_1_deg_zeros}a we show the ratio of isolated nodes
as a function of the number of iterations. 
Due to the reduced spreading in the degree
distribution with $k$ the rapid increasing tendency of $p(d=0)$ present
in the original settings is modified to a very slowly increasing tendency
for the rotated scenarios. 
The $p(d=0)$ at iteration $k=10$ is 
displayed in Fig.\ref{fig:rot_alfa_1_deg_zeros}b as a function of the 
rotation angle $\alpha$, showing an overall ``U'' shape with a minimum 
around 22.5 degrees. As pointed
out previously, when $\alpha\rightarrow 0$,
we recover the original settings of the MFNG. Although there seem to
 be many pairs of $\alpha$ values with very similar $p(d=0)$ values, we
note that each $\alpha$ defines a different setting with unique attributes
(e.g., the area of the rotated square affecting the average degree and the 
family of the $w_k(h)$ curves are always unique for each $\alpha$).
The overall properties of a rotated setting (degree distribution, ratio
of isolated nodes, etc.) change smoothly with the rotation angle for
any $\alpha>0$. However, as argued in Sect.\ref{sect:rot}. theoretically and
shall be examined in Sect.\ref{sect:Dq}. numerically, the behaviour of the
$D(q)$ curve does show a drastic change when switching from the $\alpha=0$ 
original setting to an $\alpha>0$ finite rotation angle. This change
from a multifractal $D(q)$ to a non-multifractal one is not expected
to be affected by any commensurability effect in rotation angles.

One of the big advantages of the original MFNG was that it provides a flexible
tool for generating random graphs with realistic properties. Although 
examining to what extent this feature is affected by the rotation 
of the LPM is beyond the goals of this article, we present an example where
power-law like degree distribution was achieved in the rotated scenario
in the Appendix.

\subsection{Measuring the $D(q)$ curve}
\label{sect:Dq}

The results shown in Sect.\ref{sect:deg}. are very promising, since the
rotation of the LPM drastically reduced the ratio of the isolated nodes 
in the studied example, especially in the case of larger rotation angles.
However, an even more reassuring way for checking the effect of the
 modification of the MFNG is the measurement of the $D(q)$ curves 
corresponding to the $w_k(h)$ functions. As described in 
Sect.\ref{sect:multifract}., from the $D(q)$ at $q=1$ we can deduce 
the behaviour of the fraction of isolated nodes, i.e., if $D(q=1)=1$
in the $k\rightarrow\infty$ limit, than the isolated nodes cannot become
dominant. 

Before proceeding to the $D(q)$ curves, we have one important note. 
In practise we are always dealing with multifractals at a finite
number of iterations, which have a lower size bound (lower
length bound in our case) depending on the number of iterations. 
Below this lower size bound they are not any more structured and, thus, 
this size bound provides a lower bound for the 
size $\epsilon$ of the boxes with which we cover the multifractal 
when measuring $D(q)$ as described in Sect.\ref{sect:multifract}. 

\begin{figure}[hbt]
\centerline{\includegraphics[width=0.6\textwidth]{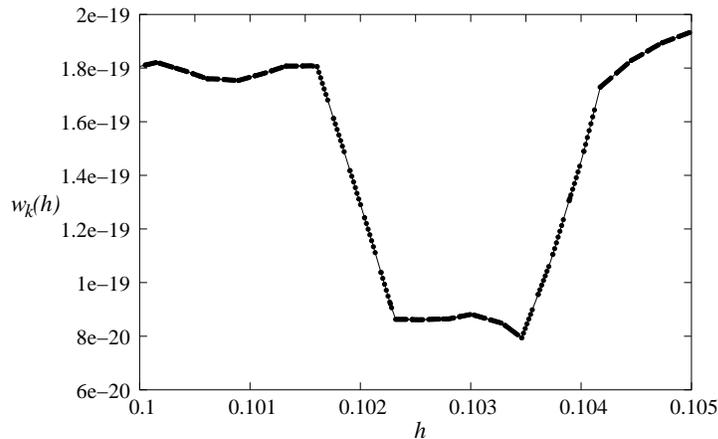}}
\caption{ A small part of the $w_k(h)$ of the rotated link probability
measure at $k=8$, $\alpha=10.5$ degrees. The points correspond to the
boundaries of the segments in which $w_k(h)$ is strictly linear.
However, for many adjacent segments the slope is actually the same or
nearly the same.
}
\label{fig:join_segms}
\end{figure}

In case of the $w_k(x)$ of the original setting (without any rotation) this 
lower bound in $\epsilon$ is given simply by the width of the 
rows (columns). In case of the $w_k(h)$ of the rotated measure
the situation is a bit more complex. First of all, when compared to the 
$w_k(x)$ of the original setting  at the same number 
of iterations, $w_k(h)$ can contain much smaller segments. 
However, there are many adjacent
segments with the same or almost the same slope of the linking probability,
which can be united into  single large segment, as
 shown in Fig.\ref{fig:join_segms}.
 Thus, before the
 application of the $D(q)$ measuring procedure above, these segment joining
were carried out, and the lower bound of $\epsilon$ was set to the average of 
the length of the resulting segments. This yielded still a much smaller
value than in case of the original measure.

\begin{figure}[hbt]
\centerline{\includegraphics[width=\textwidth]{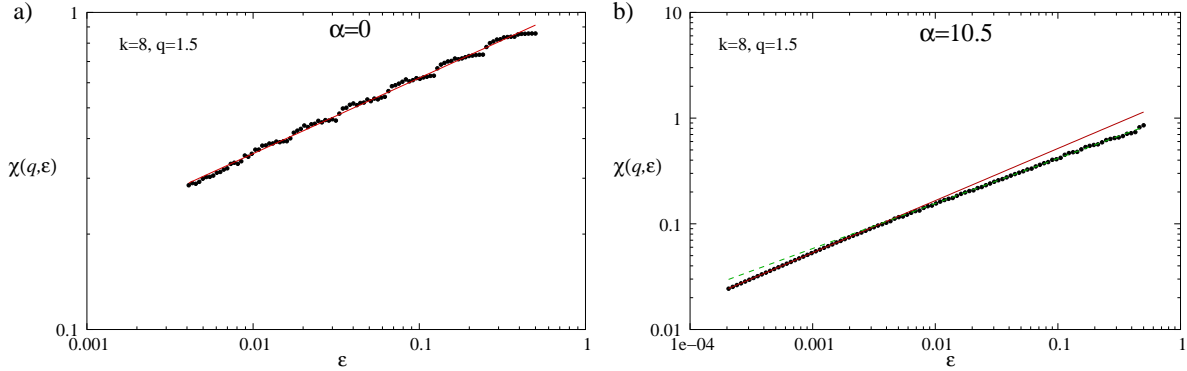}}
\caption{a) The $\chi(q,\epsilon)$ obtained for the $w_k(x)$
of the original setting at $k=8$ and $q=1.5$ on logarithmic scale 
(black symbols), together with a power-law fit (red line). b)
The $\chi(q,\epsilon)$ of the $w_k(h)$ of the rotated frame at rotation angle
$\alpha=10.5$ degrees with the same parameters $k=8$ and  $q=1.5$ on 
logarithmic scale (black symbols). In contrast to the original settings, when
fitting  $\chi(q,\epsilon)$ with a power-law, the lower part and the upper
 part of the function yield different exponents, as shown by the continuous
red line and the dashed green line respectively. }
\label{fig:chi_q_compare}
\end{figure}

Interestingly, for the rotated frames when plotting $\ln\chi(q,\epsilon)$
as a function of $\ln \epsilon$, the curves seem to consist of two 
subsequent linear segments with different slopes. In contrast, the
$\ln\chi(q,\epsilon)$ obtained from $w_k(x)$ in the rotation free settings 
shows a linear behaviour as a function of $\ln\epsilon$. The two types
of $\chi(q,\epsilon)$ curves are shown in Fig.\ref{fig:chi_q_compare}.
The $D(q)$ is obtained from the slope of these curves, which is straight
forward in case of the $w_k(x)$ of the original settings. However, which
part of the $\ln\chi(q,\epsilon)$ curve should we fit in case of the 
rotated measures? According to the definition given in (\ref{eq:def_Dq})
$D(q)$ should be evaluated in the $\epsilon\rightarrow 0$ limit, thus,
we use the slope of the lower part of the $\ln\chi(q,\epsilon)$ curves in the
 rotated scenario for measuring $D(q)$. 

\begin{figure}[hbt]
\centerline{\includegraphics[width=\textwidth]{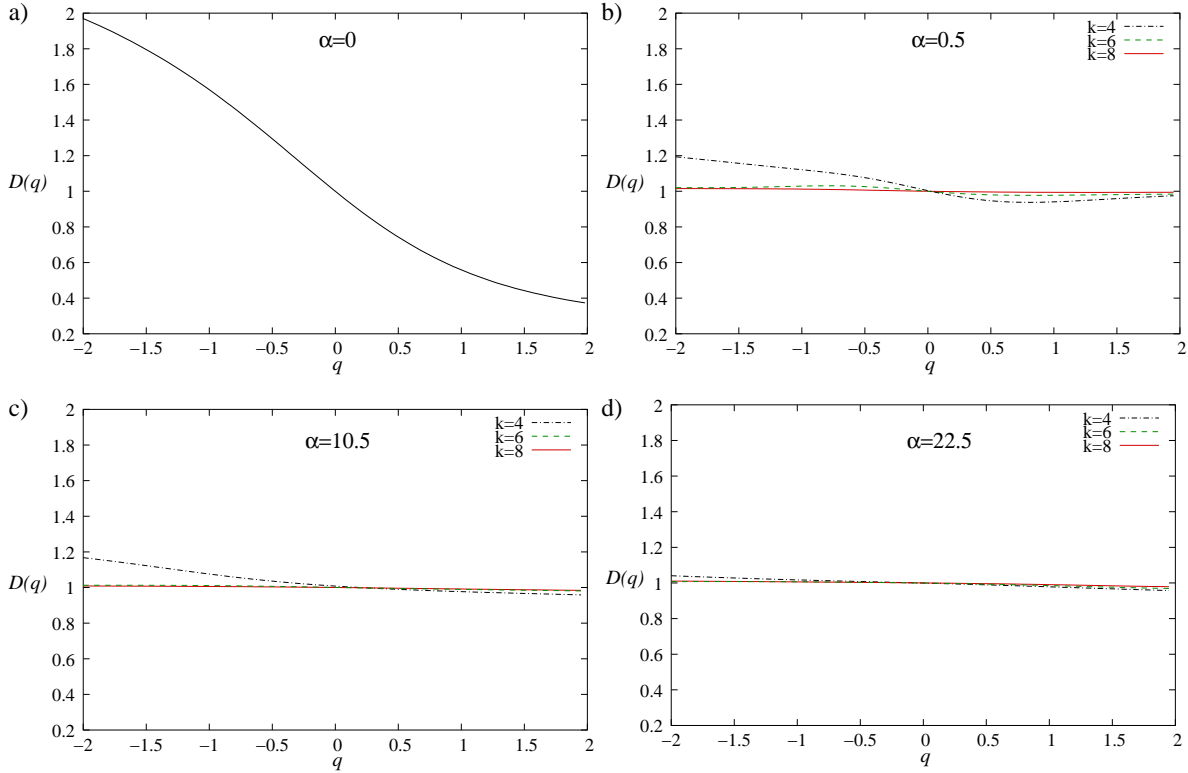}}
\caption{a) The $D(q)$ curve of the $w_k(x)$ function, corresponding to
the original, ``rotation-free'' settings, obtained from (\ref{eq:D_q_simp}). 
The $D(q)$ curves measured in the rotated scenario for $k=4,6,8$ at rotation 
angles $\alpha=0.5^{\circ}$, $\alpha=10.5^{\circ}$ and $\alpha=22.5^{\circ}$ 
are shown in panels b), c), and d), respectively.}
\label{fig:Dq_results}
\end{figure}

In Fig.\ref{fig:Dq_results}. we show the results obtained for
the $D(q)$ curves at various rotation angles. (For a comparison,
the $D(q)$ curve of the original settings is shown as well). 
From these figures it seems that $D(q)$ takes a trivial ``non-multifractal'' 
form at already $k=8$, and it is very close to 1 at $q=1$. The  
``precision'' of the numerical $D(q)$ determination process at $q=1$ 
was tested on other 1d functions where we know that $D(q=1)=1$. According
 to that, for the rotated measures the numerically obtained $D(q=1)$ 
values are equal to $D(q=1)=1$ within error bound for $k=8$.

\section{Summary}
\label{sect:sum}

In summary, we investigated the node isolation effect of the MFNG from
a new point of view. It had become clear that this phenomena is very 
closely related to the multifractality of the 1 dimensional 
projection of the LPM determining the degree distribution. According
to general theorems concerning multifractals, the projection in question 
is a particular one, and in contrast, the vast majority of the other 
1d projections of the LPM do not bear multifractal properties. 
Based on this observation we introduced a slight variation of the original
MFNG method, involving the rotation of the LPM with a given angle. 
Due to the rotation, the projection determining the degree distribution
is no longer a special projection in any aspects, thus, the node 
isolation effect is expected to disappear due to the lack of multifractality. 
The empirical studies support the theoretical reasoning above: the 
order $q$ generalised fractal dimension $D(q)$ in case of the projection of the 
LPM related to the degree
distribution became trivial, and for the numerically accessible range of the
number of iterations the fraction of isolated nodes showed 
a drastic reducement when compared to the original settings without any
rotation. 

\section*{Acknowledgment}
This work was supported
 by the Hungarian National Science Fund (OTKA K68669), the National
 Research and Technological Office (NKTH, Textrend) and the J\'anos Bolyai 
Research Scholarship of the Hungarian Academy of Sciences.

\section*{Appendix}
\subsection*{A1. Lengths}
\begin{figure}[h]
\centerline{\includegraphics[width=0.75\textwidth]{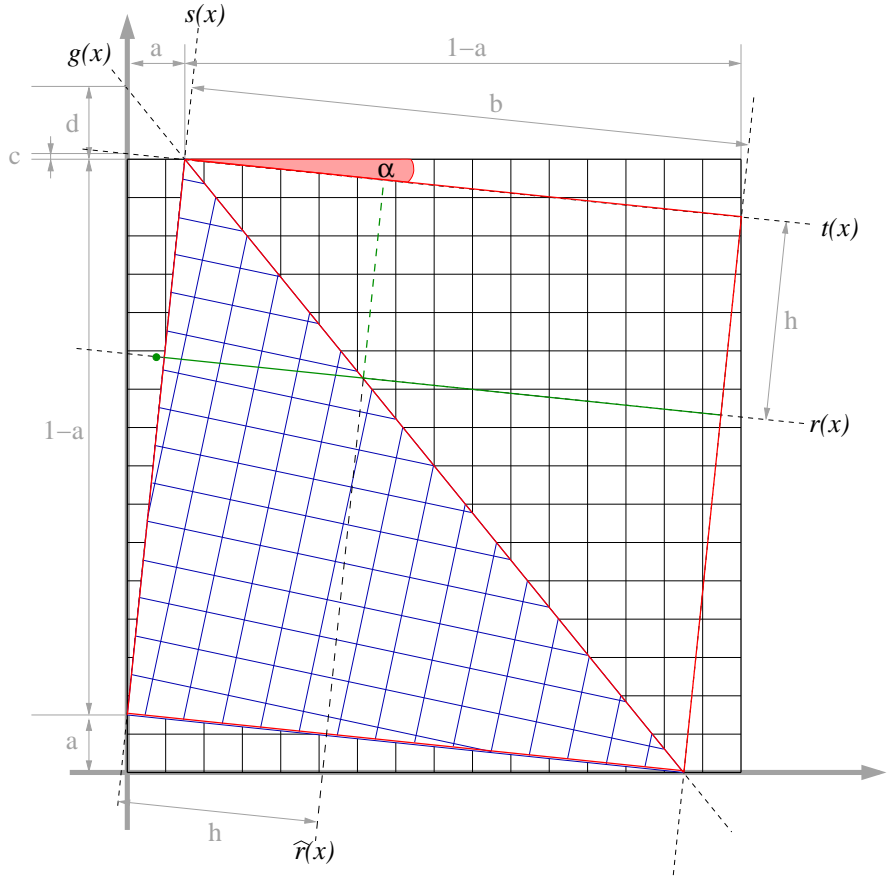}}
\caption{The geometry and the definition of the various lengths and 
sizes in the rotated link probability measure.}
\label{fig:rot_geom}
\end{figure}

For the length of the interval $a$ in Fig.\ref{fig:rot_geom} we can write 
\begin{equation}
\frac{a}{1-a}=\tan \alpha,
\end{equation}
thus,
\begin{equation}
a=\frac{\tan \alpha}{1+\tan \alpha}.
\end{equation}
Similarly, we can write for $b$
\begin{equation}
\frac{a}{b}=\sin \alpha,
\end{equation}
thus,
\begin{equation}
b=\frac{\tan \alpha}{\sin \alpha (1+\tan \alpha)}=\frac{1}{\cos \alpha+\sin \alpha}.
\end{equation}
Furthermore,
\begin{equation}
c=a\tan\alpha=\frac{\tan^2 \alpha}{1+\tan \alpha},
\end{equation}
whereas 
\begin{equation}
d=a\tan (\alpha+\pi/4)=a\frac{\sin(\alpha+\pi/4)}{\cos(\alpha+\pi/4)}=
a\frac{\sin\alpha+\cos\alpha}{\cos\alpha-\sin\alpha}.
\end{equation}

\subsection*{A2. Equations of the different lines}

In the coordinate system shown in Fig.\ref{fig:rot_geom}. the equation
of the top side of the rotated frame can be written as
\begin{equation}
t(x)=1+c-x\tan(\alpha).
\label{eq:top}
\end{equation}
Similarly, the equation of a parallel line below at a distance of
$h$ is given by
\begin{equation}
r(x)=1+c-h\cos(\alpha)-h\sin(\alpha)\tan(\alpha)-x\tan(\alpha).
\end{equation}
The equation of the diagonal can be expressed as
\begin{equation}
g(x)=1+d-x\tan(\alpha+\pi/4).
\end{equation}
Finally, the equation of the left side of the rotated square is given
by
\begin{equation}
s(x)=a+x\tan(\pi/2-\alpha),
\end{equation}
and the equation of a parallel line shifted by $h$ to the right 
(corresponding to the dashed green line in Fig.\ref{fig:rot_geom}.)
can be written as
\begin{equation}
\widehat{r}(x)=a+(x-h\cos\alpha-h\sin\alpha\tan\alpha)\tan(\pi/2-\alpha).
\end{equation}

The equations given above can be used to calculate
 where a given line intersects with a given box boundary of the 
original link-probability measure, and from the location of the 
intersection points we can calculate the lengths/areas of the 
intersections between the lines and the boxes.

\begin{figure}[hbt]
\centerline{\includegraphics[width=0.65\textwidth]{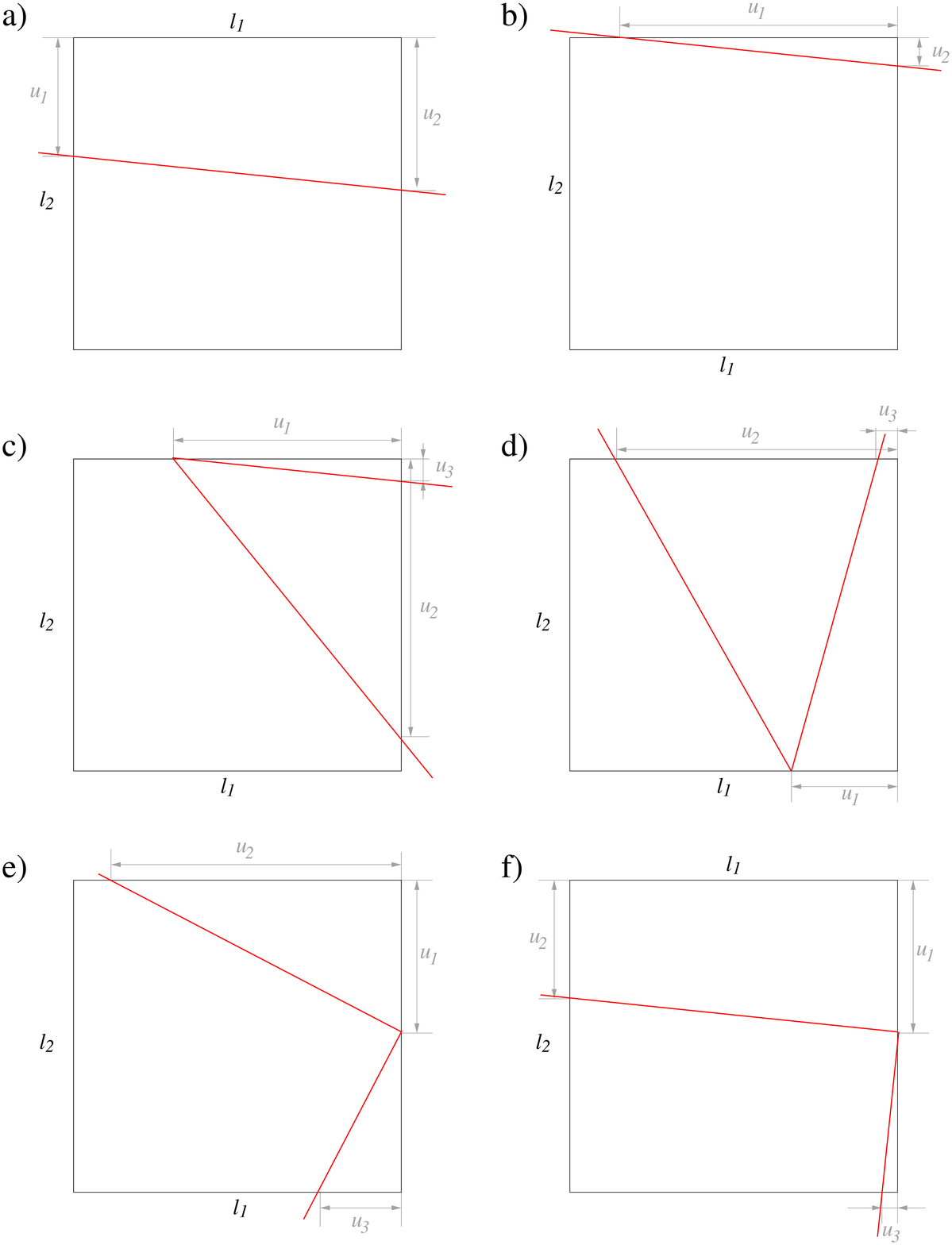}}
\caption{The six type of intersections of the boxes with the boundary
lines of the top triangular part of the rotated square.
}
\label{fig:areas}
\end{figure}

To calculate $\left< d\right>$ we have to calculate the area of each intersected
box in the top triangular part of the rotated square. The boundary lines
of this triangle can intersect with the boxes as summarised in Fig\ref{fig:areas}.:
\begin{itemize}
\item in case the box intersects with only one line, then this divides it
into either two trapezoids (Fig.\ref{fig:areas}a), or into a triangle and 
a pentagon (Fig.\ref{fig:areas}b),
\item in case the box intersects with two lines, then there are still only
three intersection points (instead of four), since the boundary lines
must intersect each other on the boundary of the box. The remaining two 
unshared intersection points can take qualitatively three different positions:
they can both fall on a side adjacent to the side of the shared intersection
point (Fig.\ref{fig:areas}c), they can both fall on the side opposite
to the shared intersection point (Fig.\ref{fig:areas}d), they can fall on
two opposite sides adjacent to the side of the shared intersection
(Fig.\ref{fig:areas}e), or
they can fall on two adjacent sides (Fig.\ref{fig:areas}f).
\end{itemize} 

\subsection*{A3. Changing coordinate system}

\begin{figure}[hbt]
\centerline{\includegraphics[width=0.65\textwidth]{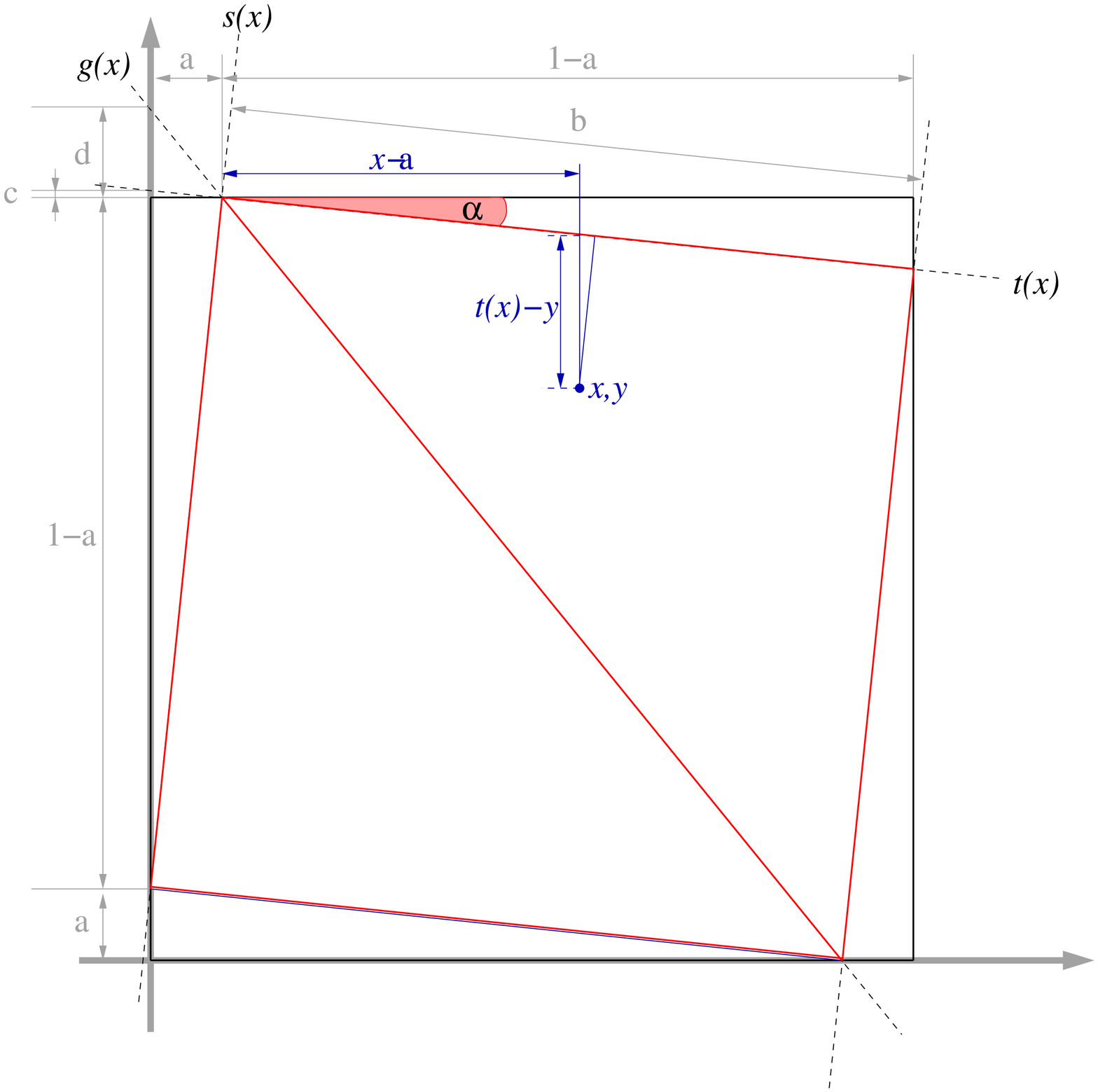}}
\caption{Changing to the coordinate system of the rotated square.}
\label{fig:coord_trans}
\end{figure}

According to Fig.\ref{fig:coord_trans}., the new
coordinates of a general point $x,y$ can be given as
\begin{eqnarray}
\tilde{x}&=&\frac{x-a}{\cos(\alpha)}+(t(x)-y)\sin(\alpha),\\
\tilde{y}&=&(t(x)-y)\cos(\alpha),
\end{eqnarray}
where $t(x)$ denotes the equation of the topside of the rotated square
in the standard coordinate system, given by (\ref{eq:top}).

\subsection*{A4. Generating skewed $\rho^{(k)}(d)$ in the rotated scenario}

\begin{figure}[hbt]
\centerline{\includegraphics[width=0.65\textwidth]{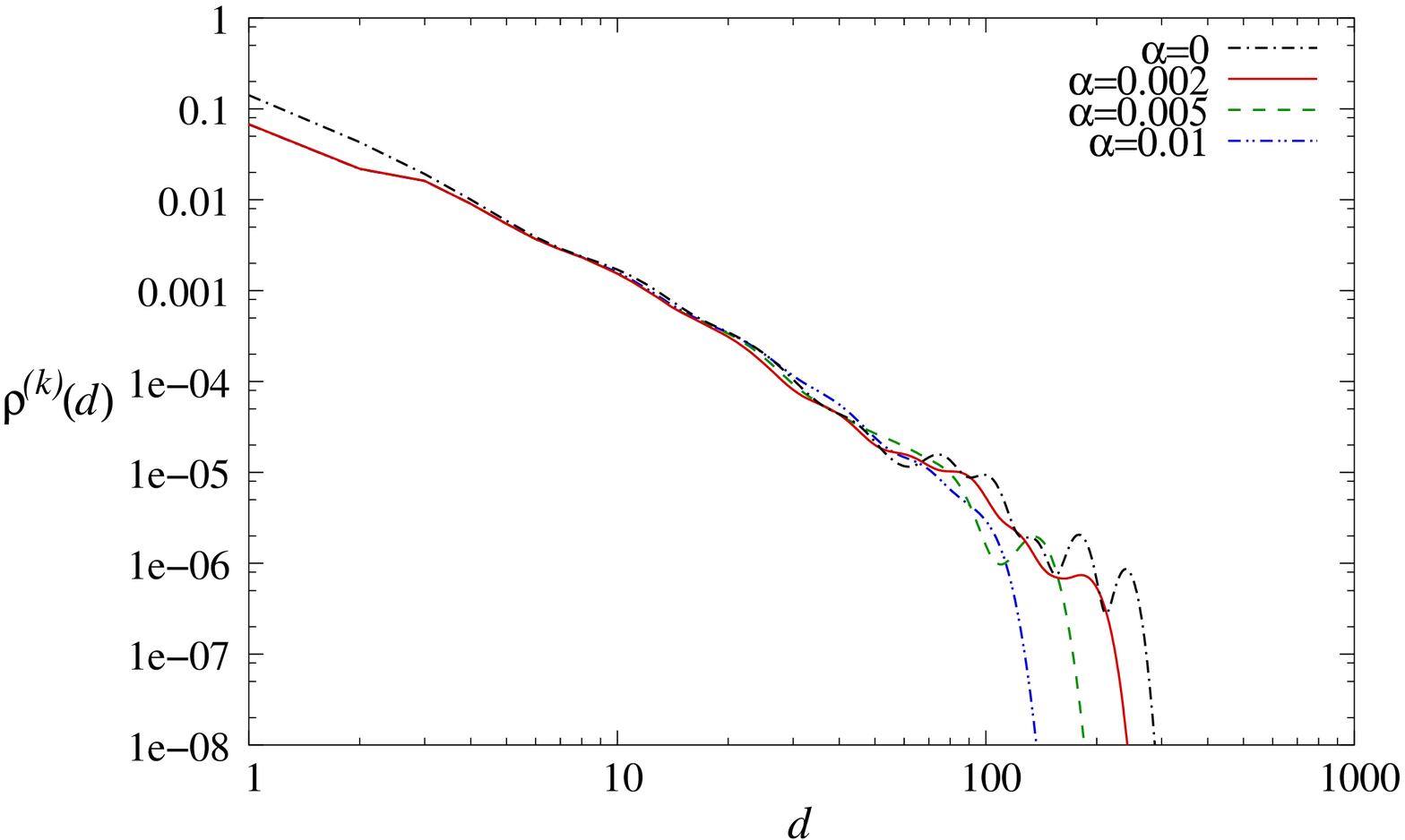}}
\caption{A power-law like degree distribution in the original settings 
at $k=4$ (black dot-dashed line), for which a slight rotation still 
preserves the skewed nature as shown by the solid 
red line corresponding to $\rho^{(k)}(d)$
at rotation angle $\alpha=0.002$ degrees. When $\alpha$ is increased, the 
skewedness of $\rho^{(k)}(d)$ is decreasing, as shown by the other two 
distributions corresponding to $\alpha=0.005$ and $\alpha=0.01$.}
\label{fig:rot_pow}
\end{figure}

In Fig.\ref{fig:rot_pow}. we show an example where power-law like 
degree distribution is generated in the rotated scenario. According to
the plots, a slight rotation preserves the degree distribution almost completely,
whereas for larger rotations the skewed nature of $\rho^{(k)}(d)$ is 
slowly disappearing.

\section*{References}

\bibliographystyle{unsrt}
\bibliography{rot_fr_njp}

\end{document}